\documentclass[letterpaper,12pt]{article}
\usepackage{tabularx} 
\usepackage{amsmath}
\usepackage{amsmath,bm}

\usepackage{graphicx} 
\usepackage[margin=1in,letterpaper]{geometry} 
\usepackage{cite} 
\usepackage[final]{hyperref} 
\hypersetup{
	colorlinks=true,       
	linkcolor=blue,        
	citecolor=blue,        
	filecolor=magenta,     
	urlcolor=blue         
}
\usepackage{amsfonts}
\usepackage{bbold}

\begin{document}
\title{Two theorems on the outer product of input and output Stokes vectors for deterministic optical systems}
\author{M. A. Kuntman and E. Kuntman}

\maketitle 
\begin{abstract}
\end{abstract}
$2\times2$ complex Jones matrix transforms two dimensional complex Jones vectors into complex Jones vectors and accounts for phase introduced by deterministic optical systems. On the other hand, Mueller-Jones matrix  transforms four parameter real Stokes vectors into four parameter real Stokes vectors
that contain no information about phase. Previously, a $4\times4$ complex matrix ($\mathbf{Z}$ matrix) was introduced. $\mathbf{Z}$ matrix is analogous to the Jones matrix and it is also akin to the Mueller-Jones matrix by the relation $\mathbf{M}=\mathbf{Z}\mathbf{Z^*}$. It was shown that $\mathbf{Z}$ matrix transforms Stokes vectors (Stokes matrices) into complex vectors (complex matrices) that contain relevant phases besides the other information. In this note it is shown that, for deterministic optical systems, there exist two relations between outer product of experimentally measured real input-output Stokes vectors and complex vectors (matrices) that represent the polarization state and phase of totally polarized output light.

\section{Introduction}

$2\times2$ complex Jones matrices represent optical properties of deterministic optical systems and two dimensional complex Jones vectors represent polarization state of totally polarized light. Jones matrices transform Jones vectors into Jones vectors. Jones matrices and Jones vectors account for phase introduced by deterministic optical systems. 

Optical properties of deterministic systems can also be represented by $4\times4$ Mueller-Jones matrices. Mueller-Jones matrices are real matrices, hence they can be obtained by polarimetric methods that rely on only intensity measurements. Mueller-Jones matrices act on four parameter real Stokes vectors and they transform real Stokes vectors into real Stokes vectors, which contain no information about phase.

Previously, a $4\times4$ complex matrix, $\mathbf{Z}$, analogous to the Jones matrix, $\mathbf{J}$, was introduced. $\mathbf{Z}$ matrix is also akin to the real Mueller-Jones matrix, $\mathbf{M}$, by the relation  $\mathbf{M}=\mathbf{Z}\mathbf{Z^*}$. It can be shown that $\mathbf{Z}$ matrices transform Stokes vectors  (Stokes matrices) into complex vectors (complex matrices) that contain relevant phase besides the other information \cite{KKA,KKPA, KKCA}. 



$\mathbf{Z}$ matrix represents optical properties of deterministic optical systems and it is defined by four dimensionless parameters $\tau, \alpha, \beta$ and $\gamma$:

\begin{equation}
\mathbf{Z}=\frac{1}{\sqrt{2}}\begin{pmatrix}
\tau&\alpha&\beta&\gamma\\
\alpha&\tau&-i\gamma&i\beta\\
\beta&i\gamma&\tau&-i\alpha\\
\gamma&-i\beta&i\alpha&\tau
\end{pmatrix}.
\end{equation}
$\tau, \alpha$, $\beta$ and $\gamma$ are generally complex numbers related with basic anisotropy parameters of the deterministic optical system \cite{KKA}. One of the parameters can always be chosen as real and positive if overall phase is not taken into account.

$\mathbf{Z}$ matrix is a mathematical object devised to act on Stokes vectors (Stokes matrices). In this note it is shown that, for deterministic optical systems, there exist two relations between outer product of real input-output Stokes vectors of totally polarized light and complex vector (matrix) states that obtained as a result of  transformation of Stokes vectors (matrices) by $\mathbf{Z}$ matrices.


\section{Relations between the matrices representing deterministic optical systems}

$\mathbf{Z}$ matrix is a $4\times4$ version of the Jones matrix, $\mathbf{J}$. The relation between $\mathbf{Z}$ and $\mathbf{J}$ matrices can be disclosed by writing the Jones matrix in terms of parameters $\tau, \alpha, \beta$ and $\gamma$:

\begin{equation}\mathbf{J}=\frac{1}{\sqrt{2}}\begin{pmatrix}
\tau+\alpha&\beta-i\gamma\\ \beta+i\gamma&\tau-\alpha
\end{pmatrix}
\end{equation}

$\mathbf{Z}$ matrix is also closely related with the real Mueller-Jones matrix, $\mathbf{M}$. In terms of $\mathbf{Z}$ matrices, Mueller matrix of nondepolarizing optical system can be written as,
\begin{equation}\label{ZZ=M}
\mathbf{M}= \mathbf{Z}\mathbf{Z^*}=\mathbf{Z^*}\mathbf{Z}.
\end{equation}
Eq.\eqref{ZZ=M} leads to an expression for the Mueller-Jones matrix in terms of basic parameters 
$\tau, \alpha, \beta$ and $\gamma$ \cite{KKA}:
\begin{equation}\label{table2}
\mathbf{M}=\frac{1}{2}\left(\begin{array}{c|c|c|c}
\tau\tau^*+\alpha\alpha^*&\tau\alpha^*+\alpha\tau^*&\tau\beta^*+\beta\tau^*&\tau\gamma^*+\gamma\tau^*\\
\beta\beta^*+\gamma\gamma^*&+i(\gamma\beta^*-\beta\gamma^*)&+i(\alpha\gamma^*-\gamma\alpha^*)&+i(\beta\alpha^*-\alpha\beta^*)\\
\hline
\tau\alpha^*+\alpha\tau^*&\tau\tau^*+\alpha\alpha^*&\alpha\beta^*+\beta\alpha^*&\alpha\gamma^*+\gamma\alpha^*\\
-i(\gamma\beta^*-\beta\gamma^*)&-\beta\beta^*-\gamma\gamma^*&+i(\tau\gamma^*-\gamma\tau^*)&+i(\beta\tau^*-\tau\beta^*)\\
\hline
\tau\beta^*+\beta\tau^*&\alpha\beta^*+\beta\alpha^*&\tau\tau^*-\alpha\alpha^*&\beta\gamma^*+\gamma\beta^*\\
-i(\alpha\gamma^*-\gamma\alpha^*)&-i(\tau\gamma^*-\gamma\tau^*)&+\beta\beta^*-\gamma\gamma^*&+i(\tau\alpha^*-\alpha\tau^*)\\
\hline
\tau\gamma^*+\gamma\tau^*&\alpha\gamma^*+\gamma\alpha^*&\beta\gamma^*+\gamma\beta^*&\tau\tau^*-\alpha\alpha^*\\
-i(\beta\alpha^*-\alpha\beta^*)&-i(\beta\tau^*-\tau\beta^*)&-i(\tau\alpha^*-\alpha\tau^*)&-\beta\beta^*+\gamma\gamma^*\\
\end{array}\right).
\end{equation}

From Eq.\eqref{table2} , by direct calculation, it can be shown that $tr(\mathbf{M}\mathbf{M}^T)=4M_{00}^2$.


\section{Transformation of polarization states of light by the states representing the optical system}

Mueller matrices transform  four parameter real Stokes vectors $|s\rangle$ into four parameter real Stokes vectors $|s'\rangle$:
\begin{equation}
|s'\rangle=\mathbf{M}|s\rangle,
\end{equation}
where $|s\rangle=(s_0, s_1, s_2, s_3)^T$,  $|s'\rangle= (s'_0,s'_1,s'_2,s'_3)^T$; $s_i, s'_i$ are real numbers.
If $s_0^2=s_1^2+s_2^2+s_3^2$ the light is totally polarized.
If the Mueller matrix is nondepolarizing (optical system is deterministic) and if $|s\rangle$ represents totally polarized light then $|s'\rangle$ is also totally polarized, $(s'_0)^2=(s'_1)^2+(s'_2)^2+(s'_3)^2$.  

On the other hand, $\mathbf{Z}$ matrices transform real Stokes vectors of totally polarized light, $|s\rangle$, into complex vectors, $|\Tilde{s}\rangle$:
\begin{equation}\label{s=Zs}
|\Tilde{s}\rangle=\mathbf{Z}|s\rangle,
\end{equation}
where $|\Tilde{s}\rangle=(\Tilde{s}_0,\Tilde{s}_1,\Tilde{s}_2,\Tilde{s}_3)^T$ and $\Tilde{s}_0, \Tilde{s}_1, \Tilde{s}_2$ and $\Tilde{s}_3$ are, in general, complex numbers.

$\mathbf{Z}$ matrix also transforms $\mathbf{S}$ matrices (Stokes matrices) into $\mathbf{\Tilde{S}}$ matrices:

\begin{equation}\label{S=ZS}
\mathbf{\Tilde{S}}=\mathbf{Z}\mathbf{S}\end{equation}
where $\mathbf{S}$ and $\mathbf{\Tilde{S}}$ are matrix states representing input and output polarization states of totally polarized light, and they are defined as follows:

\begin{equation}\mathbf{S}=\frac{1}{\sqrt{2}}\begin{pmatrix}s_0&s_1&s_2&s_3\\s_1&s_0&-is_3&is_2\\s_2&is_3&s_0&-is_1\\s_3&-is_2&is_1&s_0\end{pmatrix}\end{equation}

\begin{equation}\mathbf{\Tilde{S}}=\frac{1}{\sqrt{2}}\begin{pmatrix}\Tilde{s}_0&\Tilde{s}_1&\Tilde{s}_2&\Tilde{s}_3\\\Tilde{s}_1&\Tilde{s}_0&-i\Tilde{s}_3&i\Tilde{s}_2\\\Tilde{s}_2&i\Tilde{s}_3&\Tilde{s}_0&-i\Tilde{s}_1\\\Tilde{s}_3&-i\Tilde{s}_2&i\Tilde{s}_1&\Tilde{s}_0\end{pmatrix}\end{equation}
$\mathbf{S}$ matrix is a complex matrix but it is based on real Stokes parameters, hence, it may be appropriate to name it as "Stokes matrix". $\mathbf{\Tilde{S}}$ matrix is a complex matrix based on complex parameters, $\Tilde{s}_i$, and it represents the
polarization state and phase
of totally polarized output
light. There is a connection between  $\mathbf{\Tilde{S}}$ matrix and $|\Tilde{s}\rangle$ vector: First column of $\mathbf{\Tilde{S}}$ matrix is proportional to $|\Tilde{s}\rangle$ vector, because, Eq.\eqref{s=Zs} is closely related with Eq.\eqref{S=ZS} and with their quaternion counterparts.

It can be shown that, just like complex Jones matrices, $\mathbf{Z}$ matrices and hence $|\Tilde{s}\rangle$ vectors (and $\mathbf{\Tilde{S}}$ matrices) bear total phases introduced by optical systems on totally polarized input light\cite{KKCA}:
\begin{equation}
\langle E|E'\rangle=\langle E|\mathbf{J}|E\rangle=\frac{1}{\sqrt{2}}\langle s|\mathbf{Z}|s\rangle=\frac{1}{\sqrt{2}}\langle s|\Tilde{s}\rangle.
\end{equation}
where $\mathbf{J}$ is the Jones matrix, $|E\rangle$ is the Jones vector, and $|E'\rangle=\mathbf{J}|E\rangle$.
\section{Matrices defined by outer products of vector states}

A real matrix $\mathbf{Q}$ is defined as an outer product of $|s\rangle$ vectors:

\begin{equation}
\mathbf{Q}=|s\rangle\langle s|.
\end{equation}

A real matrix $\mathbf{K}$ is defined as an outer product of $|s'\rangle$ and $|s\rangle$ vectors:

\begin{equation}
\mathbf{K}=|s'\rangle\langle s|,
\end{equation}
or,
\begin{equation}
\mathbf{K}=\mathbf{M}\mathbf{Q}.
\end{equation}
If $|s\rangle$ and $|s'\rangle$ represent totally polarized light, then it can be shown that \begin{equation}
tr(\mathbf{K}\mathbf{K}^T)=4K_{00}^2.
\end{equation}

A complex-Hermitian matrix $\mathbf{\mathfrak{S}}$ is defined as an outer product of $|\Tilde{s}\rangle$  vector with its Hermitian conjugate:

\begin{equation}
\mathbf{\mathfrak{S}}=|\Tilde{s}\rangle\langle \Tilde{s}|.
\end{equation}



\section{Two theorems for outer product of input and output Stokes vectors}

In this note it is shown that there exists two relations between real Stokes vectors ($|s\rangle, |s'\rangle$) and complex vectors and matrices ($|\Tilde{s}\rangle, \mathbf{\Tilde{S}}$) representing polarization state and phase of totally polarized output light. 

First relation is between $\mathbf{K}=|s'\rangle\langle s|$ matrix and matrix product of $\mathbf{\Tilde{S}}$ with its complex conjugate:

\begin{equation}
\mathbf{K}=|s'\rangle\langle s|=\mathbf{\Tilde{S}}\mathbf{\Tilde{S}}^*.
\end{equation}
Proof is straightforward. From $\mathbf{M}=\mathbf{Z}\mathbf{Z}^*$, Mueller-Jones matrix, $\mathbf{M}$, is written in terms of parameters $\tau, \alpha, \beta$ and $\gamma$. From $\mathbf{M}|s\rangle=|s'\rangle$, $\mathbf{K}$ can be obtained in terms of $\tau, \alpha, \beta$ and $\gamma$. Then, it can be shown that $\mathbf{K}_{i,j}=(\mathbf{\Tilde{S}}\mathbf{\Tilde{S}}^*)_{i,j}$.

Second relation can be formulated between real $\mathbf{K}$ and complex $\mathbf{\mathfrak{S}}$ matrix ($\mathbf{\mathfrak{S}}=|\Tilde{s}\rangle\langle
\Tilde{s}|$). $|s'\rangle$ and $|s\rangle$ vectors are real and directly measurable quantities and $\mathbf{K}$ has the following explicit form:
\begin{equation}
\mathbf{K}=
\begin{pmatrix}s'_0s_0& s'_0s_1&s'_0s_2&s'_0s_3\\s'_1s_0&s'_1s_1&s'_1s_2&s'_1s_3\\s'_2s_0&s'_2s_1&s'_2s_2&s'_2s_3&\\s'_3s_0&s'_3s_1&s'_3s_2&s'_3s_3
\end{pmatrix}.
\end{equation}
On the other hand, $\mathbf{\mathfrak{S}}$ matrix is a complex-Hermitian matrix that contains information about phase introduced by the optical system which cannot be measured by simple polarimetric methods and it has the following explicit form:
\begin{equation}
\mathbf{\mathfrak{S}}=\begin{pmatrix}\Tilde{s}_0\Tilde{s}^*_0&\Tilde{s}_0\Tilde{s}^*_1&\Tilde{s}_0\Tilde{s}^*_2&\Tilde{s}_0\Tilde{s}^*_3\\\Tilde{s}_1\Tilde{s}^*_0&\Tilde{s}_1\Tilde{s}^*_1&\Tilde{s}_1\Tilde{s}^*_2&\Tilde{s}_1\Tilde{s}^*_3\\ \Tilde{s}_2\Tilde{s}^*_0&\Tilde{s}_2\Tilde{s}^*_1&\Tilde{s}_2\Tilde{s}^*_2&\Tilde{s}_2\Tilde{s}^*_3&\\ \Tilde{s}_3\Tilde{s}^*_0&\Tilde{s}_3\Tilde{s}^*_1&\Tilde{s}_3\Tilde{s}^*_2&\Tilde{s}_3\Tilde{s}^*_3
\end{pmatrix}.
\end{equation}

It can be shown that $\mathbf{K}$  and $\mathbf{\mathfrak{S}}$ matrices can be bridged by a $\mathbf{\mathbb{\Sigma}}$ transformation:

\begin{equation}
\mathbf{\mathfrak{S}}=\mathbf{\mathbb{\Sigma}}(\mathbf{K}),
\end{equation}
where the $\mathbf{\mathbb{\Sigma}}$ transformation is defined as

\begin{equation}\label{sigmaK}
\mathbf{\mathbb{\Sigma}}(\mathbf{K})=\frac{1}{2}\sum_{i,j=0}^3 K_{ij}{\Sigma}_{ij},\end{equation}
or, in an explicit form:
{\footnotesize
\begin{equation}\label{table}\mathbf{\mathbb{\Sigma}}(\mathbf{K})=\frac{1}{2}\left(\begin{array}{c|c|c|c}
K_{00}+K_{11}&K_{01}+K_{10}&K_{02}+K_{20}&K_{03}+K_{30}\\
K_{22}+K_{33}&-i(K_{23}-K_{32})&+i(K_{13}-K_{31})&-i(K_{12}-K_{21})\\
\hline
K_{01}+K_{10}&K_{00}+K_{11}&K_{12}+K_{21}&K_{13}+K_{31}\\
+i(K_{23}-K_{32})&-K_{22}-K_{33}&+i(K_{03}-K_{30})&-i(K_{02}-K_{20})\\
\hline
K_{02}+K_{20}&K_{12}+K_{21}&K_{00}-K_{11}&K_{23}+K_{32}\\
-i(K_{13}-K_{31})&-i(K_{03}-K_{30})&+K_{22}-K_{33}&+i(K_{01}-K_{10})\\
\hline
K_{03}+K_{30}&K_{13}+K_{31}&K_{23}+K_{32}&K_{00}-K_{11}\\
+i(K_{12}-K_{21})&+i(K_{02}-K_{20})&-i(K_{01}-K_{10})&-K_{22}+K_{33}\\
\end{array}\right)
\end{equation}}
Details of the transformation can be found in the Appendix.

The relation $\mathbf{\mathfrak{S}}=\mathbf{\mathbb{\Sigma}}(\mathbf{K})$ can be proved by calculating each element of  $\mathbf{\mathbb{\Sigma}}(\mathbf{K})$ to show that $\mathfrak{S}_{i,j}=(\mathbf{\mathbb{\Sigma}}(\mathbf{K}))_{i,j}$. 


\section{An example}
As an example, let $\tau=1+i,\: \alpha=1-2i,\: \beta=2+3i$ and $\gamma=0$.
Corresponding $\mathbf{Z}$ matrix is
\begin{equation}
\mathbf{Z}=\frac{1}{\sqrt{2}}\begin{pmatrix}
1+i&1-2i&2+3i&0\\1-2i&1+i&0&-3+2i\\2+3i&0&1+i&-2-i\\0&3-2i&2+i&1+i
\end{pmatrix}
\end{equation}

The Mueller matrix of the deterministic optical system can be easily obtained from the relation $\mathbf{M}=\mathbf{Z}\mathbf{Z}^*$:
\begin{equation}
\mathbf{M}=\frac{1}{2}\begin{pmatrix}
20&-2&10&-14\\
-2&-6&-8&-2\\
10&-8&10&-6\\
14&2&6&-16
\end{pmatrix}
\end{equation}
For a given input Stokes vector of totally polarized light all relevant vectors and matrices can be calculated.  For example let $s_0=5, s_1=3, s_2=0, s_3=4$ be four real Stokes parameters representing the polarization state of totally polarized light:

\begin{equation}|s\rangle=\begin{pmatrix}
5\\3\\0\\4
\end{pmatrix},\quad
|s'\rangle=\mathbf{M}|s\rangle=\begin{pmatrix}
19\\-18\\1\\6
\end{pmatrix};\quad |\Tilde{s}\rangle=\mathbf{Z}|s\rangle=\frac{1}{\sqrt{2}}\begin{pmatrix}
8-i\\-4+i\\2+11i\\13-2i
\end{pmatrix}.
\end{equation}

\begin{equation}
\mathbf{K}=|s'\rangle\langle s|=\begin{pmatrix}95&57&0&76\\-90&-54&0&-72\\5&3&0&4\\30&18&0&24
\end{pmatrix}
\end{equation}

$\mathbf{\Tilde{S}}$ matrix can be written as follows
\begin{equation}
\mathbf{\Tilde{S}}=\frac{1}{2}\begin{pmatrix}
8-i&-4+i&2+11i&13-2i\\
-4+i&8-i&-2-13i&-11+2i\\
2+11i&2+13i&8-i&1+4i\\
13-2i&11-2i&-1-4i&8-i
\end{pmatrix}
\end{equation}
By direct multiplication it can be shown that $\mathbf{K}=\mathbf{\Tilde{S}}\mathbf{\Tilde{S}}^*
$.

In order to show the second relation, $\mathbf{K}$ matrix is subjected to a $\mathbf{\mathbb{\Sigma}}$ transformation given by Eq.\eqref{sigmaK}:

\begin{equation}
\mathbf{\mathbb{\Sigma}}(\mathbf{K})=\frac{1}{2}\sum_{i,j=0}^3 K_{ij}{\Sigma}_{ij}=\frac{1}{2}\begin{pmatrix}65&-33-4i&5-90i&106+3i\\-33+4i&17&3+46i&-54+5i\\5+90i&3-46i&125&4+147i\\106-3i&-54-5i&4-147i&173
\end{pmatrix}.
\end{equation}
It is now easy to show $\mathbf{\mathfrak{S}}=\mathbf{\mathbb{\Sigma}}(\mathbf{K})$ by calculating the outer product $\mathbf{\mathfrak{S}}=|\Tilde{s}\rangle\langle \Tilde{s}|$:
\begin{equation}
\mathbf{\mathfrak{S}}=\frac{1}{2}\begin{pmatrix}8-i\\-4+i\\2+11i\\13-2i
\end{pmatrix}\begin{pmatrix}8+i,&-4-i,&2-11i,&13+2i
\end{pmatrix}.
\end{equation}

It is worth noting that $rank(\mathbf{\mathfrak{S}})=1$, hence, all column vectors of $\mathbf{\mathfrak{S}}$ matrix are equivalent to each other, i.e., they differ from each other only by respective phases, and they are also
equivalent to the $|\Tilde{s}\rangle$ vector apart from overall phase factors. For example the first column vector of $\mathbf{\mathbb{\Sigma}}(\mathbf{K})$ matrix, $|c_1\rangle$, differs from the $|\Tilde{s}\rangle$ vector by a factor $(8+i)/\sqrt{2}$:

\begin{equation}
|c_1\rangle=\frac{1}{2}\begin{pmatrix}
65\\-33+4i\\5+90i\\106-3i
\end{pmatrix}=\frac{8+i}{\sqrt{2}}|\Tilde{s}\rangle .
\end{equation}

Now, suppose that $\mathbf{Z}$ matrix and the Mueller 
matrix $\mathbf{M}$ are not given, but $|s\rangle$ and $|s'\rangle$ vectors are known as
a result of measurement, then $|\Tilde{s}\rangle$ vector can be calculated from the outer product of $|s\rangle$ and $|s'\rangle$ vectors apart from its original overall phase. 
Situation is very similar to the overall phase issue encountered while trying to calculate the Jones matrix of the optical system from the 
associated nondepolarizing Mueller matrix \cite{arxiv1}.

Complex components of $|\Tilde{s}\rangle$ vector are not easily accessible by measurement, but $|\Tilde{s}\rangle$ vector plays a very important role in the extended and unified formalism of polarization optics \cite{zenframework}.
If $|\Tilde{s}\rangle$ is given or calculated it is possible to extract corresponding $|s\rangle$ and
$|s'\rangle$ vectors from $\mathbf{\mathfrak{S}}=|\Tilde{s}\rangle\langle \Tilde{s}|$ by the following inverse transformation:

\begin{equation}\label{K}
\mathbf{K}=\mathbf{\mathbb{\Sigma}}(\mathbf{\mathfrak{S}})=\frac{1}{2}\sum_{i,j=0}^3 \mathfrak{S}_{ij}{\Sigma}_{ij}.
\end{equation}
Once $\mathbf{K}$ matrix is calculated from $\mathbf{\mathfrak{S}}$ matrix, it can be shown that $|s'\rangle$ and $|s\rangle$ vectors (input and output Stokes vectors) can be obtained from $\mathbf{K}=|s'\rangle\langle s|$. In the above example, $|s\rangle$ vector can be read from the first row of $\mathbf{K}$ matrix and $|s'\rangle$ can be read from the first column of $\mathbf{K}$ matrix:

\begin{equation}
\textbf{(First row)}^T=\begin{pmatrix}95\\57\\0\\76\end{pmatrix}=s'_0|s\rangle=19\begin{pmatrix}5\\3\\0\\4
\end{pmatrix}
\end{equation}

\begin{equation}
\textbf{First column}=\begin{pmatrix}95\\-90\\5\\30\end{pmatrix}=s_0|s'\rangle=5\begin{pmatrix}19\\-18\\1\\6
\end{pmatrix}
\end{equation}
In order to recover the original $|s\rangle$ and $|s'\rangle$ vectors, actual values of the parameters $s'_0$ and $s_0$ are needed. But, usually, absolute values of the input and output Stokes parameters are not important, hence, $|s\rangle$ and $|s'\rangle$ vectors  can be re-normalized for further calculations.


\section{Conclusion}

Jones matrix transforms two dimensional complex Jones vectors into complex Jones vectors and accounts for the phase introduced by deterministic optical systems. Mueller-Jones matrix of deterministic optical system transforms real Stokes vectors into real Stokes vectors which contains no information about phase. A $4\times4$ complex matrix $\mathbf{Z}$ matrix transforms Stokes vectors and Stokes matrices into  complex vectors $|\Tilde{s}\rangle$ and complex matrices $\mathbf{\Tilde{S}}$ that contain relevant phase besides the other information. In this note it is shown that, for deterministic optical systems, there exists two relations between outer product of real input-output Stokes vectors and  complex vectors/matrices that represent polarization state and phase of totally polarized output light.

Matrix $\mathbf{K}$ is defined as an outer product of real Stokes vectors $|s\rangle$ and $|s'\rangle$: $\mathbf{K}=|s'\rangle\langle s|$, where $|s'\rangle=\mathbf{M}|s\rangle$. $\mathbf{Z}$ matrix acts of the real Stokes vector: $\mathbf{Z}|s\rangle=|\Tilde{s}\rangle$, where $|\Tilde{s}\rangle$ is a complex valued vector that can account for the phase. $\mathbf{Z}$ matrix also transforms Stokes matrices: $\mathbf{Z}\mathbf{S}=\mathbf{\Tilde{S}}$, where $\mathbf{\Tilde{S}}$ is a complex valued matrix with relevant phase. 
First relation is that $\mathbf{K}=\mathbf{S}\mathbf{S}^*$, and the second relation can be written as $\mathbf{\mathbb{\Sigma}}(\mathbf{K)=\mathbf{\mathfrak{S}}}$, where $\mathbf{\mathfrak{S}}=|\Tilde{s}\rangle\langle\Tilde{s}|$; $\mathbf{\mathbb{\Sigma}}$ is a special transformation from a real matrix to a complex-Hermitian matrix.

These relations show that $\mathbf{K}$ matrix may have importance in the mathematical framework of polarization algebra. $\mathbf{K}$ matrix serves as an interface where the real-measurable parameters meet complex parameters which are rather remote from the experiment \cite{zenframework}. It can be shown that K matrix may also have practical implications.
For example, it can be shown that, by using either one of the theorems involving K matrix, the Jones matrix, and hence the Mueller-Jones matrix of a deterministic optical system with certain symmetry properties, can be obtained from the results of two polarimetric measurements. \cite{zenJbytwo}.


\section{Appendix}
Let $\mathbf{G}$ be any $4\times4$ real matrix. $\mathbf{G}$ can be transformed into an associated complex-Hermitian matrix $\mathbf{\mathfrak{G}}$ by the transformation $\mathbb{\Sigma}$\cite{KKA}:

\begin{equation}\label{Transform}
\mathbf{\mathfrak{G}}=\mathbb{\Sigma}(\mathbf{G})= \frac{1}{2}\sum_{i,j=0}^3G_{ij}{\Sigma}_{ij},
\end{equation}
where $G_{ij} (i,j = 0,1,2,3)$ are the elements of the real matrix $\mathbf{G}$, and
${\Sigma}_{ij}=\mathbf{U}({\sigma}_{i}\otimes{\sigma}_{j}^*)\mathbf{U}^{-1}$ with,

{\footnotesize\begin{equation}\mathbf{U}=\frac{1}{\sqrt{2}}\begin{pmatrix}1&0&0&1\\1&0&0&-1\\0&1&1&0\\0&i&-i&0\end{pmatrix},\quad\mathbf{U}^{-1}=\mathbf{U}^{\dagger}=\frac{1}{\sqrt{2}}\begin{pmatrix}1&1&0&0\\0&0&1&-i\\0&0&1&i\\1&-1&0&0\end{pmatrix}.\end{equation}}
The superscript $^{\dagger}$ indicates the complex conjugate and transpose, the superscript $^*$ indicates complex conjugate, $\otimes$ is the Kronecker product and ${\sigma_{i}}$ are the Pauli matrices with the $2\times2$ identity in the following order:
\begin{equation}{\sigma}_0=\begin{pmatrix}1&0\\0&1\end{pmatrix},\quad{\sigma}_1=\begin{pmatrix}1&0\\0&-1\end{pmatrix}\quad{\sigma}_2=\begin{pmatrix}0&1\\1&0\end{pmatrix},\quad{\sigma}_3=\begin{pmatrix}0&-i\\i&0\end{pmatrix}.\end{equation}

Explicit form of the $\mathbf{\mathfrak{G}}$ matrix can be obtained from the following transformation table:
{\footnotesize
\begin{equation}\label{Transformation}
\mathbf{\mathfrak{G}}=\frac{1}{2}\left(\begin{array}{c|c|c|c}
G_{00}+G_{11}&G_{01}+G_{10}&G_{02}+G_{20}&G_{03}+G_{30}\\
G_{22}+G_{33}&-i(G_{23}-G_{32})&+i(G_{13}-G_{31})&-i(G_{12}-G_{21})\\
\hline
G_{01}+G_{10}&G_{00}+G_{11}&G_{12}+G_{21}&G_{13}+G_{31}\\
+i(G_{23}-G_{32})&-G_{22}-G_{33}&+i(G_{03}-G_{30})&-i(G_{02}-G_{20})\\
\hline
G_{02}+G_{20}&G_{12}+G_{21}&G_{00}-G_{11}&G_{23}+G_{32}\\
-i(G_{13}-G_{31})&-i(G_{03}-G_{30})&+G_{22}-G_{33}&+i(G_{01}-G_{10})\\
\hline
G_{03}+G_{30}&G_{13}+G_{31}&G_{23}+G_{32}&G_{00}-G_{11}\\
+i(G_{12}-G_{21})&+i(G_{02}-G_{20})&-i(G_{01}-G_{10})&-G_{22}+G_{33}\\
\end{array}\right)
\end{equation}}
It is worth noting that rank of the Hermitian matrix $\mathbf{\mathfrak{G}}$ can take any value between $1-4$, in general.  

Eq. \eqref{Transform} can be inverted by the same transformantion, $\mathbb{\Sigma}$:

\begin{equation}\label{Inverse}
\mathbf{G}=\mathbb{\Sigma}(\mathbf{\mathfrak{G}})= \frac{1}{2}\sum_{i,j=0}^3\mathfrak{G}_{ij}{\Sigma}_{ij},
\end{equation}

 


\end{document}